\PassOptionsToPackage{table,dvipsnames}{xcolor}
\documentclass[conference,10pt]{IEEEtran}
\IEEEoverridecommandlockouts

\newcommand{\egl}[1]{{\color{blue}\bf{[EGL: #1]}}}

\renewcommand{\egl}[1]{}

\newcommand{\hoang}[1]{{\color{ForestGreen} \bf KHN: #1}}

\renewcommand{\hoang}[1]{}

\newcommand{\revise}[1]{{\color{red}#1}}

\renewcommand{\revise}[1]{{#1}}

\usepackage[left=0.625in,right=0.625in,top=0.705in,bottom=1.03in]{geometry}
\usepackage{cite}
\usepackage{amsmath,amssymb,amsfonts,amsthm}
\usepackage{algorithmic}
\usepackage{graphicx}
\usepackage{textcomp}
\usepackage{xcolor}
\usepackage{enumitem}
\usepackage{mathtools}
\usepackage{comment}
\PassOptionsToPackage{hyphens}{url}\usepackage[hidelinks]{hyperref}
\usepackage{tikz,pgfplots}
\usepackage{grffile}
\usepgfplotslibrary{patchplots}
\usetikzlibrary{backgrounds,plotmarks,fit,positioning,arrows,shapes,shapes.multipart,calc,arrows.meta}
\usetikzlibrary{fit,positioning,arrows,shapes,shapes.multipart,calc,arrows.meta,shapes.geometric,shapes.misc}
\usetikzlibrary{decorations.pathreplacing,angles,quotes,calligraphy}
\usetikzlibrary{automata}

\allowdisplaybreaks

\usepackage[labelformat=simple]{subcaption}

\usepackage{vmr-symbols-vecbold-RanDetSame}
\usepackage{standard-macros}

\include{acronym}

\newcommand{\tc}{\tau\sub{c}}
\newcommand{\tpl}{\tau\sub{p}}
\newcommand{\tu}{\tau\sub{u}}
\newcommand{\td}{\tau\sub{d}}

\newcommand{\ts}{\tau\sub{s}}
\newcommand{\tg}{\tau\sub{g}}

\newcommand{\DS}{{\sf DS}}
\newcommand{\BU}{{\sf BU}}
\newcommand{\UI}{{\sf UI}}

\newcommand{\SEdl}{{\sf SE}}

\newcommand{\Pdl}{\rho\sub{AP}}
\newcommand{\Ppl}{\rho\sub{UE}}

\def\BibTeX{{\rm B\kern-.05em{\sc i\kern-.025em b}\kern-.08em T\kern-.1667em\lower.7ex\hbox{E}\kern-.125emX}}
    
\begin{document}

\title{Breaking the TDD Flow for \revise{Over-the-Air} Phase Synchronization in Distributed Antenna Systems
    \vspace{-.3cm}
}

\author{
\IEEEauthorblockN{Khac-Hoang Ngo and Erik G. Larsson}

 \IEEEauthorblockA{Department of Electrical Engineering (ISY), Linköping University, Linköping, SE-58183 Sweden} \vspace{-.6cm}
 \thanks{This work was supported by ELLIIT,  the Swedish Research Council (VR),
    and the KAW foundation.}
}

\maketitle
\begin{abstract}
    Phase synchronization between distributed antenna arrays requires measurements that break the standard time-division duplex (TDD) operation. We present a feasibility study on implementing such synchronization and analyze its impact on the quality of service. Considering two antenna arrays with independent local oscillators (LOs), we propose a modified TDD flow to accommodate the transmission of phase synchronization signals, formulate the phase estimation and compensation problem, and derive the achievable downlink spectral efficiency (SE). Numerical results show that frequent re-estimation of the inter-array phase disparity is essential for maximizing SE in systems with low-quality LOs. Furthermore, applying a Kalman filter for phase tracking substantially improves the SE, \revise{especially if phase estimation errors are large compared to LOs phase drifts.
    }
\end{abstract}

\section{Introduction} \label{sec:intro}

Distributed antenna architectures~\cite{Xu25} are considered a key technology component of the 6G physical layer. In a distributed antenna system, \glspl{AP}\textemdash each with a single or multiple antennas\textemdash are spread out geographically and cooperate phase-coherently to serve \glspl{UE}. The \gls{TDD} flow allows the system to exploit channel reciprocity and perform downlink beamforming based on uplink channel estimates. A major challenge arises when the \glspl{AP} are driven by different \glspl{LO} that are not locked in phase. In this case, phase synchronization\footnote{Also called phase calibration, and phase alignment,
in the literature.} measurements are required for joint reciprocity calibration across  \glspl{AP}~\cite{Rogalin14,Kim22,Rashid23,Larsson23_calibration,Larsson24_synchrony,Ganesan24_beamsync}. This calibration must be repeated whenever the \glspl{LO} have drifted significantly. 

Phase synchronization measurements disrupt the \gls{TDD} flow. Specifically, over a part of a slot assigned for downlink transmission, if an \gls{AP} transmits a phase synchronization signal, the other \glspl{AP} must switch from downlink to uplink mode in order to receive this signal. This switch cannot occur instantly, as it must respect the guard intervals between uplink and downlink transmissions. This ``broken'' \gls{TDD} flow introduces a nontrivial tradeoff. On the one hand, increasing the uplink/downlink overlap allows more frequent measurements for phase tracking. On the other hand, this prevents the \gls{AP} that switched to uplink mode from sending data to the \glspl{UE},   thus potentially reduces the \gls{SE}. To date, this \gls{TDD}-breaking mechanism and the mentioned tradeoff have not been formally analyzed.

In this paper, we formulate a framework for breaking the conventional \gls{TDD} flow to enable phase synchronization between distributed \glspl{AP}. We consider a system with two \glspl{AP} and propose a mechanism where one of the \glspl{AP} periodically shifts its uplink and downlink periods in certain \gls{TDD} slots, such that its downlink period partially overlaps with the uplink period of the other \gls{AP}, and vice versa
(see Fig.~\ref{fig:TDD_operation}, to be described in detail later). During these overlapping time, the downlink \gls{AP} transmits a phase synchronization signal to the uplink \gls{AP}, resulting, over time, in bi-directional measurements. This allows for reciprocity calibration in the next period where both \glspl{AP} are in downlink. To enhance the phase estimation accuracy, we also develop a Kalman filter for phase tracking. We then derive the downlink \gls{SE} achieved with conjugate beamforming after phase synchronization.
In numerical experiments, we investigate the optimal frequency of phase synchronization, represented by the number of slots over which the phase is re-estimated once, to maximize the \gls{SE}. 
\revise{We found that frequent re-estimation of the inter-array phase disparity is essential for maximizing \gls{SE} in systems with low-quality LOs.
Additionally, the Kalman filter provides significantly gain, especially when the \glspl{LO} have high quality and the inter-\gls{AP} link is weak, so that phase estimation errors are large compared to \glspl{LO} drifts.}
\begin{figure}[t!]
    \centering
    \newlength{\UL}
    \setlength{\UL}{1.5cm}

    \newlength{\DL}
    \setlength{\DL}{2cm}

    \newlength{\guard}
    \setlength{\guard}{.3cm}

    \newlength{\sync}
    \setlength{\sync}{.2cm}

    \newlength{\pilot}
    \setlength{\pilot}{.25cm}
    
    \tikzset{uplink/.style={
    draw,rectangle,
    minimum height=.7cm,
    minimum width=\UL
    }}
    \tikzset{downlink/.style={
    draw,rectangle,
    minimum height=.7cm,
    minimum width=\DL
    }}

    \tikzset{rectangle open right/.style={
            draw=none, minimum height=.65cm,
            append after command={
                [shorten <= -0.5\pgflinewidth]
                ([shift={(-1.5\pgflinewidth,-0.5\pgflinewidth)}]\tikzlastnode.north east)
            edge([shift={( 0.5\pgflinewidth,-0.5\pgflinewidth)}]\tikzlastnode.north west) 
                ([shift={( 0.5\pgflinewidth,-0.5\pgflinewidth)}]\tikzlastnode.north west)
            edge([shift={( 0.5\pgflinewidth,+0.5\pgflinewidth)}]\tikzlastnode.south west)            
                ([shift={( 0.5\pgflinewidth,+0.5\pgflinewidth)}]\tikzlastnode.south west)
            edge([shift={(-1.0\pgflinewidth,+0.5\pgflinewidth)}]\tikzlastnode.south east)
            }
        }
    }
    
    \tikzset{rectangle open left/.style={
            draw=none, minimum height=.65cm,
            append after command={
                (\tikzlastnode.north west)
            edge(\tikzlastnode.north east) 
                (\tikzlastnode.north east)
            edge(\tikzlastnode.south east)            
                (\tikzlastnode.south east)
            edge(\tikzlastnode.south west)
            }
        }
    }

    \captionsetup[subfigure]{oneside,margin={0cm,0cm}}
    \subcaptionbox{Conventional TDD flow. Blue boxes represent uplink pilot transmissions.}{\scalebox{.95}{\scalebox{.83}{\begin{tikzpicture}
    \node[uplink] (ul11) at (0,0) {uplink};
    \node[rectangle open left, left=\guard of ul11] (l1) {$\!\ldots $};
    \node[left=0cm of l1] (AP1) {$\!$AP $1$};
    \node[downlink,right=\guard of ul11] (dl11) {downlink};
    \node[uplink,right=\guard of dl11] (ul12) {uplink};
    \node[downlink,right=\guard of ul12] (dl12) {downlink};
    \node[rectangle open right, right=\guard of dl12] (r1) {$ \ldots\!$};

    \node[uplink] (ul21) at (0,-1cm) {uplink};
    \node[rectangle open left, left=\guard of ul21] (l2) {$\!\ldots $};
    \node[left=0cm of l2] (AP2) {$\!$AP~$2$};
    \node[downlink,right=\guard of ul21] (dl21) {downlink};
    \node[uplink,right=\guard of dl21] (ul22) {uplink};
    \node[downlink,right=\guard of ul22] (dl22) {downlink};
    \node[rectangle open right, right=\guard of dl22] () {$ \ldots\!$};

    \draw[gray!50] ([yshift=.5cm]ul11.north west) -- ([yshift=-1.2cm]ul11.south west);
    \draw[gray!50] ([yshift=.5cm]ul12.north west) -- ([yshift=-1.2cm]ul12.south west);
    \draw[gray!50] ([yshift=.5cm]r1.north west) -- ([yshift=-1.2cm]r1.south west);

    \fill[blue!70,opacity=.5] (ul11.north west) rectangle ([xshift=\pilot]ul11.south west);
    \fill[blue!70,opacity=.5] (ul21.north west) rectangle ([xshift=\pilot]ul21.south west);
    \fill[blue!70,opacity=.5] (ul12.north west) rectangle ([xshift=\pilot]ul12.south west);
    \fill[blue!70,opacity=.5] (ul22.north west) rectangle ([xshift=\pilot]ul22.south west);
    
    \draw[<->] ([yshift=.3cm]ul11.north west) -- node[midway,above] () {$\tc$} ([yshift=.3cm]ul12.north west);
\end{tikzpicture}
}} \label{fig:TDD}} \\
    \captionsetup[subfigure]{oneside,margin={0cm,0cm}}
    \subcaptionbox{Proposed ``broken'' TDD flow: once every $F$ slots, \gls{AP}~2 moves the last $\ts + \tg$ samples of the uplink to the end of the slot, and shifts the downlink earlier accordingly. Orange boxes represent the instants of phase measurements, with arrows showing the direction of synchronization signals.
    }{\scalebox{.95}{\scalebox{.83}{\begin{tikzpicture}
    \node[uplink] (ul11) at (0,0) {uplink};
    \node[rectangle open left, left=\guard of ul11] (l1) {$\!\ldots$};
    \node[left=0cm of l1] (AP1) {$\!$AP~$1$};
    \node[downlink,right=\guard of ul11] (dl11) {downlink};
    \node[uplink,right=\guard of dl11] (ul12) {uplink};
    \node[downlink,right=\guard of ul12] (dl12) {downlink};
    \node[rectangle open right, right=\guard of dl12] (r1) {$\ldots\!$};

    \node[uplink,minimum width=\UL-\guard-\sync] (ul21) at (-0.5*\guard-0.5*\sync,-1cm) {$\!$uplink$\!$};
    \node[rectangle open left, left=\guard of ul21] (l2) {$\!\ldots$};
    \node[left=0cm of l2] (AP2) {$\!$AP~$2$};
    \node[downlink,right=\guard of ul21] (dl21) {downlink};
    \node[uplink,minimum width=\UL+\guard+\sync,right=\guard of dl21] (ul22) {uplink};
    \node[downlink,right=\guard of ul22] (dl22) {downlink};
    \node[rectangle open right, right=\guard of dl22] () {$\ldots\!$};

    \draw[gray!70] ([yshift=.5cm]ul11.north west) -- ([yshift=-1.2cm]ul11.south west);
    \draw[gray!70] ([yshift=.5cm]ul12.north west) -- ([yshift=-1.2cm]ul12.south west);
    \draw[gray!70] ([yshift=.5cm]r1.north west) -- ([yshift=-1.2cm]r1.south west);

    \fill[blue!80,opacity=.5] (ul11.north west) rectangle ([xshift=\pilot]ul11.south west);
    \fill[blue!80,opacity=.5] (ul21.north west) rectangle ([xshift=\pilot]ul21.south west);
    \fill[blue!80,opacity=.5] (ul12.north west) rectangle ([xshift=\pilot]ul12.south west);
    \fill[blue!80,opacity=.5] ([xshift=\guard+\sync]ul22.north west) rectangle ([xshift=\pilot+\guard+\sync]ul22.south west);
    
    \fill[orange] (dl11.south east) rectangle ([xshift=-\sync]dl11.north east);

    \fill[orange] (ul11.south east) rectangle ([xshift=-\sync]ul11.north east);

    \fill[orange] (dl21.south west) rectangle ([xshift=\sync]dl21.north west);
    \fill[orange] (ul22.south west) rectangle ([xshift=\sync]ul22.north west);

    \draw[-latex] ([xshift=-0.5*\sync,yshift=.35cm]dl11.south east) -- ([xshift=0.5*\sync,yshift=.35cm]ul22.south west);

    \draw[-latex] ([xshift=0.5*\sync,yshift=.35cm]dl21.south west) -- ([xshift=-0.5*\sync,yshift=.35cm]ul11.south east);

    \draw[->] ([xshift=0.5*\sync,yshift=-.35cm]dl21.south west) -- node[below=.1cm,near start,align = center] () {time $i_1$, \\ measure \\ $\nu_{1,i_1}- \nu_{2,i_1}$} ([xshift=0.5*\sync,yshift=-.05cm]dl21.south west);

    \draw[->]([xshift=0.5*\sync,yshift=-.35cm]ul22.south west) -- node[below=.1cm,near start,align = center] () {time $i_2$, \\  measure \\ $~~~\nu_{2,i_2} - \nu_{1,i_2}$} ([xshift=0.5*\sync,yshift=-.05cm]ul22.south west);

    \draw[<->] ([yshift=.3cm]ul11.north west) -- node[midway,above] () {$\tc$} ([yshift=.3cm]ul12.north west);
\end{tikzpicture}
}}\label{fig:shifted_TDD}}
    \vspace{-.13cm}
    \caption{Illustration of two consecutive slots following the conventional or proposed ``broken'' \gls{TDD} operation. 
    } 
    \label{fig:TDD_operation}
    \vspace{-.65cm}
\end{figure}

\subsubsection*{Notation}
	We denote scalars with plain italic letters, 
    column vectors with lowercase boldface letters, 
    and matrices with uppercase boldface letters. 
    The superscripts~$^*$, $\tp{}$, and~$\herm{}$ denote the conjugate, transpose, and conjugate transpose, respectively.
	By $\matidentity_m$ and $\mathbf{0}_m$, we denote the $m\times m$ 
    identity matrix and $m\times 1$ all-zero vector, respectively. 
    We denote the set of integers from~$1$ to $n$ by $[n]$, and the imaginary unit by $\jmath$.
    
\section{System Model} \label{sec:model}

We consider a system with two \glspl{AP}, each having $N$ antennas, serving $K$ single-antenna \glspl{UE}. The \glspl{AP} are connected to a \gls{CPU} via fronthaul links. Each \gls{AP} is driven by a single, independent~\gls{LO}. 

\subsection{Channel and Phase Noise Model}
We consider a narrowband channel where samples are taken
sequentially in time without inter-symbol interference. We also refer to ``sample'' as a discretized time unit. Let $\vech_{k,\ell} \in \complexset^{N}$ denotes the channel vector between \gls{UE}~$k\in [K]$ and \gls{AP}~$\ell\in \{1,2\}$. 
We consider a block fading channel model where $\vech_{k,\ell}$ remains constant for each \revise{time} block of $\tc$ samples and varies independently between blocks. 
Furthermore, we consider \gls{iid} Rayleigh fading with $\vech_{k,\ell} \sim \jpg(\mathbf{0}, \beta_{k,\ell}\matidentity_{N})$, where $\beta_{k,\ell}$ is the large-scale fading coefficient. 
Let $\matG \in \complexset^{N \times N}$ be the channel from \gls{AP}~1 to \gls{AP}~2. We assume that the \glspl{AP} knows $\matG$. This assumption
is reasonable when the APs are static and thus their slowly
varying channels can be tracked accurately. 



We assume that phase noise at the  \glspl{UE} is perfectly tracked with sufficiently frequent demodulation pilots, which we do not model. We focus on phase noise at the \glspl{AP}. The free-running \gls{LO} at \gls{AP}~$\ell$ induces a phase noise $\nu_{\ell,i}$, common to all antennas, at sample time~$i$. 
The resulting multiplicative transmit and receive phase noise terms are $\exp(-\jmath \nu_{\ell,i})$ and $\exp(\jmath \nu_{\ell,i})$, respectively. Here, following the convention in~\cite{Larsson23_calibration,Larsson24_synchrony}, and consistent with~\cite{Nissel22}, we use opposite signs for the transmit and receive phase noise, as transmitted and received signals travel in opposite directions. Furthermore, we have assumed that the slowly-varying imbalance between the transmit and receive chains is negligible, and thus both transmit and receive phase noise are driven by the same random process $\nu_{\ell,i}$. 

\subsection{Signal Model}
We consider \gls{TDD} operation. Each \gls{TDD} slot corresponds to a length-$\tc$ coherence block, and is divided into different periods: 1) $\tpl$ samples for uplink pilot transmission, 2) $\tu$ samples for uplink data transmission, 3) $\td$ samples for downlink data transmission, and 4) two guard periods of $\tg$ samples each that separate the uplink and downlink periods. During both uplink pilot transmission and uplink data transmission, we say that the \gls{AP} is in uplink mode. In this paper, we focus on the analysis of the downlink data transmission.  Note that LO phase noise
is not a problem for the uplink, since on uplink pilots and
data see the same channel. We next describe the uplink pilot transmission and downlink data transmission periods without specifying their positions within a slot. 

\subsubsection{Uplink Pilot Transmission}
We assume that $\tpl = K$ and let 
user~$k$ transmit \revise{the known pilot sequence $\sqrt{\Ppl K} \vece_k$, where $\vece_k$ is the length-$K$ canonical basis vector with a one at position~$k$ and zeros elsewhere.} %
The received signal over the pilot period correlated with $\vece_k$ is equal to  the received signal at \revise{sample} time $k$ of the current slot, and is given by
\egl{consider: instead of "impulse signal" (what does that mean)something like transmit a known pilot sequence, and then after the receiver correlates with that pilot we get...}
\hoang{I use this explicit pilot sequence to simplify so that the received correlated pilot signal contains phase noise at time $k$ only} 
\begin{equation}
    \vecy_{k,\ell}\supp{pilot} = \sqrt{\Ppl K} \exp(\jmath \nu_{\ell,k}) \vech_{k,\ell} + \vecz\supp{pilot}_{k,\ell}
\end{equation}
where $\vecz\supp{pilot}_{k,\ell} \sim \jpg(\mathbf{0}_{N}, \matidentity_{N})$ is the \gls{AWGN}. 
The subscript $k$ in $\nu_{\ell,k}$ refers to sample time $k$ of the current slot. As $\nu_{\ell,k}$ is uniform over $[-\pi,\pi]$, only the effective channel $\vecq_{k,\ell} = \exp(\jmath \nu_{\ell,k}) \vech_{k,\ell}$ can be estimated. Its \gls{LMMSE} estimate is 
\begin{equation}
    \hat \vecq_{k,\ell} = c_{k,\ell} \vecy_{k,\ell}\supp{pilot} 
\end{equation}
with $c_{k,\ell} = \dfrac{\sqrt{\Ppl K} \beta_{k,\ell}}{\Ppl K \beta_{k,\ell} +  1}$. It holds that $\hat \vecq_{k,\ell}  \sim  \jpg \left(\mathbf{0}_{N}, \gamma_{k,\ell}\matidentity_{N}\right)$  with $\gamma_{k,\ell} = \sqrt{\Ppl K} \beta_{k,\ell} c_{k,\ell}$. 


\subsubsection{Downlink Data Transmission}
Consider downlink data transmission at sample time~$i$. Here, we let the index $i$ be incremented across slots, i.e., $i$ can be larger than $\tc$.  We allow for the possibility that only one \gls{AP} is in downlink mode at a given time, and let $a_{\ell,i} \in \{0,1\}$ indicate if AP~$\ell$ is in downlink period ($a_{\ell,i} = 1$) or not ($a_{\ell,i} = 0$) at time~$i$. Specifically, using conjugate beamforming, \gls{AP}~$\ell$ transmits  
    \egl{shall we consider zero-forcing multiplexing (not in this paper necessarily)?}
    \hoang{Yes, we shall do this, but in the extended version due to the time constraint.}
\begin{equation}
    \vecx_{\ell,i} = a_{\ell,i} e^{\jmath \revise{\theta_{\ell,i}}} \sqrt{\Pdl} \sum_{k=1}^K \sqrt{\frac{\eta_{k,\ell}}{N \gamma_{k,\ell}}} \hat \vecq_{k,\ell,i}^* s_{k,i},
\end{equation}
where 
$s_{k,i}$, with $\Exop[|s_{k,i}|^2] = 1$, is the data signal intended for \gls{UE}~$k$; 
$\hat \vecq_{k,\ell,i}$ is the latest estimate of $\vecq_{k,\ell}$ up to time $i$; $\eta_{k,\ell}$, $\ell \in \{1,2\}$, $k \in [K]$, are power control coefficients to satisfy the power constraint $\Exop[\|\vecx_{\ell,i}\|^2] \le \Pdl$.
The power constraint can be rewritten as 
    $\sum_{k=1}^K \eta_{k,\ell} \le 1, \ell \in \{1,2\}$. 
\gls{UE}~$k$ receives
\begin{align}
        y_{k,i} &= \sum_{\ell=1}^2 \exp({-\jmath \nu_{\ell,i}}) \tp{\vech}_{k,\ell} \vecx_{\ell,i} + z_{k,i} \\
        &= \sqrt{\Pdl}\sum_{\ell = 1}^2 a_{\ell,i} \exp(-\jmath -\nu_{\ell,i}) 
        \tp{\vech}_{k,\ell} \notag \\
        &\quad \cdot \sum_{k'=1}^K \sqrt{\frac{\eta_{k',\ell}}{N\gamma_{k',\ell}}}\hat\vecq_{k',\ell,i}^* s_{k',i} + z_{k,i} \\
        &= \sqrt{\Pdl}\sum_{\ell = 1}^2 a_{\ell,i} \sum_{k'=1}^K\sqrt{\frac{\eta_{k',\ell}}{N\gamma_{k',\ell}}} \exp[-\jmath (\nu_{\ell,i} -\nu_{\ell,[i]_k}) 
        ] \notag \\
        &\quad \cdot \tp{\vecq}_{k,\ell,i} \hat\vecq_{k',\ell,i}^* s_{k',i}  + z_{k,i} \label{eq:signal_dl} 
    \end{align}
where $z_{k,i} \sim \jpg(0,1)$ is the \gls{AWGN}. In~\eqref{eq:signal_dl}, we used that the effective channel estimated by $\hat \vecq_{k,\ell,i}$ is $\vecq_{k,\ell,i} = \exp(\jmath \nu_{\ell, [i]_k}) \vech_{k,\ell}$ where
\begin{equation}
    [i]_k = i - 1 - [(i - 1 - k) \mod \tc]    
\end{equation}
is the time index when $\hat \vecq_{k,\ell,i}$ is obtained. The index $[i]_k$ simply refers to the $k$th sample of the slot containing $i$. Because of the phase drift between the time when the channel is estimated and when it is used, both $\nu_{\ell,i}$ and $\nu_{\ell,[i]_k}$ appear in the phase. 
    In the absence of phase drift, $\nu_{\ell,i} = \nu_{\ell,[i]_k}$, as in~\cite{Larsson23_calibration,Larsson24_synchrony}.
    

\vspace{-.05cm}
\subsection{Phase Noise Compensation} \label{sec:PN_compensation}

In the received signal $ y_{k,i}$, 
the desired signal $s_{k,i}$ is corrupted by the phase noise $-\nu_{\ell,i} - \nu_{\ell,[i]_k}$. To reduce complexity, we ignore the difference between $\{[i]_k\}_{k\in [K]}$ and represent all $\{\nu_{\ell,i} + \nu_{\ell,[i]_k}\}_{k \in [K]}$ by 
\begin{equation}
    \phi\of{\ell}_i = 
    \nu_{\ell,i} +\nu_{\ell,[i]_{\floor{K/2}}}. \label{eq:omega_ell}
\end{equation} 
To compensate for this phase noise, we let \gls{AP}~$\ell$ multiply its downlink signal $\vecx_{\ell,i}$ by $\exp(\jmath \theta_{\ell,i})$ 
with
\begin{equation}
     \theta_{2,i} =  
     \phi\of{2}_i  - \phi\of{1}_i, \quad \theta_{1,i} = 0. 
     \label{eq:manyAP_theta}
\end{equation}
\gls{UE}~$k$ sees approximately the phase shift $- \phi\of{1}_i$, 
and compensates for it by estimating it via a downlink demodulation pilot. As phase estimation via demodulation pilot is well understood, we omit its details. 
Next, we propose a scheme for the \glspl{AP} to estimate $\phi\of{2}_i - \phi\of{1}_i$ 
via bidirectional measurements.

\vspace{-.05cm}
\section{Breaking TDD for Phase Synchronization}\label{sec:shifted_TDD}
Our strategy is to estimate $\phi\of{2}_i - \phi\of{1}_i$ 
at periodic times $i$, and reset the phase compensation term $\theta_{\ell,i'}$, 
$\forall i' > i$, 
as in~\eqref{eq:manyAP_theta} 
whenever a new estimate is obtained.\footnote{In phase noise processes with independent increments, such as the Wiener process, the \gls{MMSE} predictor of future values is given by the current estimate, as the future is independent of the past.}
To estimate $\phi\of{2}_i - \phi\of{1}_i$, 
we modify the TDD flow to accommodate the transmission of phase synchronization signals as follows. Starting with the conventional \gls{TDD} flow shown in Fig.~\ref{fig:TDD_operation}(a), we group $F$ slots into a frame and let \gls{AP}~2 shift the periods within the first slot of each frame 
as illustrated in Fig.~\ref{fig:TDD_operation}(b). Specifically, AP~2 moves the last $\tg+1$ samples of the uplink period to the end of the slot, and shifts the downlink earlier accordingly. This way, the uplink of AP~1 has one overlapping sample with the downlink of AP~2, and vice versa. During these overlapping samples, the downlink AP can transmit a phase synchronization signal to the uplink AP. 
This broken TDD pattern is communicated to the \glspl{UE} with negligible cost.



\subsection{Phase Measurement} \label{sec:phase_measurement}
The indices of samples where phase synchronization signals are sent in the current slot are 
\begin{equation}
    i_1 = \tpl + \tu, \quad i_2 = \tpl + \tu + \tg + \td,
\end{equation}
as indicated in Fig.~\ref{fig:TDD_operation}(b).
At time $i_1$, \gls{AP}~$1$ is in uplink, and \gls{AP}~$2$ transmits the synchronization signal
    $\vecx\supp{sync}_{2} = 
    \sqrt{\Pdl} \vecu_2$
where $\vecu_2$ is a unit-norm vector.
\gls{AP}~$1$ receives
\begin{align}
    \vecy_{1,i_1}\supp{sync} &= \sqrt{\Pdl}\exp\big(\jmath \alpha\of{2 \to 1}_{i_1}\big) \matG 
    \vecu_2 
    + \vecz_{1,i_1}\supp{sync} 
\end{align}
where $\alpha\of{2 \to 1}_{i_1} =  -\nu_{2,i_1} + \nu_{1,i_1}$. \gls{AP}~$1$ estimates $\alpha\of{2 \to 1}_{i_1}$ by
$\bar \alpha\of{2 \to 1}_{i_1} = \angle \herm{\vecu}_2 \herm{\matG} \vecy_{1,i_1}\supp{sync}$. Similarly, at time $i_2$, \gls{AP}~$1$ transmits the synchronization signal
    $\vecx\supp{sync}_{1} = 
    \sqrt{\Pdl} \vecu_1$ where $\vecu_1$ is a unit-norm vector.
\gls{AP}~$2$ receives
\begin{equation}
    \vecy_{2,i_2}\supp{sync} = \sqrt{\Pdl}\exp\big(\jmath \alpha\of{1 \to 2}_{i_2}\big) \tp{\matG} \vecu_1 
     + \vecz_{2,i_2}\supp{sync} 
\end{equation}
with $\alpha\of{1 \to 2}_{i_2} =  -\nu_{1,i_2} + \nu_{2,i_2}$, and estimates $\alpha\of{1 \to 2}_{i_2}$ by
$\bar \alpha\of{1 \to 2}_{i_2} = \angle \herm{\vecu}_1 \matG^* \vecy_{2,i_2}\supp{sync}$. During times $i_1$ and $i_2$, we assume that no data is transmitted. The \glspl{AP} send $\bar \alpha\of{2\to 1}_{i_1}$ and $\bar \alpha\of{1\to 2}_{i_2}$ to the \gls{CPU}, which 
uses
\begin{equation}
    \bar \alpha_{i_2} = \bar \alpha\of{1\to 2}_{i_2} - \bar \alpha\of{2\to 1}_{i_1}  \label{eq:2AP_phi_hat} 
\end{equation}
as an estimate of $(\nu_{2,i_1} + \nu_{2,i_2}) - (\nu_{1,i_1} + \nu_{1,i_2})$. This method also works when the synchronization signals are transmitted without beamforming, i.e., for single-antenna \glspl{AP}.
%

Following an analysis similar to~\cite[Sec.~III-B]{Ganesan24_beamsync}, we set $\vecu_1$ and~$\vecu_2$ as the leading left- and leading right-singular vectors of~$\matG$, respectively. 
The \gls{MSE} of the estimation of both $\alpha\of{2\to 1}_{i_1}$ and $\alpha\of{1\to 2}_{i_2}$ can be approximated as $\frac12 \Pdl^{-1}\opnorm{\matG}^{-2}$, where $\opnorm{\matG}$ is the operator norm 
of $\matG$. 

    We remark that the two APs only need to know $\matG$ up to a common phase shift, because 
    this phase shift is present in both $\alpha\of{2\to 1}_{i_1}$ and $\alpha\of{1\to 2}_{i_2}$, and thus canceled in their difference.  See~\cite[Appendix]{Ganesan24_beamsync} for a detailed explanation.
While the \gls{CPU} can directly use 
$\bar \alpha_{i_2}$
as an estimate of $\phi\of{2}_{i_2} - \phi\of{1}_{i_2}$ (ignoring the time offset between $i_1$ and $[i_2]_{\floor{K/2}}$), 
 the estimation of $\phi\of{2}_i - \phi\of{1}_i$ can be improved using a Kalman filter that we present next. 

\subsection{Kalman Filter} \label{sec:kalman}
To make the design of the Kalman filter explicit, we consider the discrete-time Wiener phase noise process
\begin{align}
    \nu_{\ell,i} = \nu_{\ell,i-1} + \delta_{\ell,i}, \label{eq:PN_Wiener}
\end{align}
with $\delta_{\ell,i}$ \gls{iid} as $\normal(0,\sigma_\nu^2)$ across \revise{sample} time $i$. We further assume that the \glspl{AP} have identical \gls{LO} quality represented by 
$\sigma_\nu^2 = 4 \pi^2 f\sub{c}^2 c_\nu/ f\sub{s}$ where $f\sub{c}$ is the carrier frequency, $c_\nu$ is a given constant, 
and \revise{$f\sub{s}$ is the signal bandwidth}. 

Let the process $\{\alpha_n\}_{n = 1,2,\dots}$ collect the values of $\phi\of{2}_i - \phi\of{1}_i$ at instances $i_2$ of consecutive slots where phase estimation is performed. 
The process evolution of $\alpha_n$ is
\begin{equation}
    \alpha_{n} = \alpha_{n-1} + \zeta_{n} \label{eq:kalman_process}
\end{equation}
where the noise $\{\zeta_{n}\}$ are \gls{iid} as $\normal(0,\sigma_\zeta^2)$ with $\sigma_\zeta^2 = [8F\tau_c - 4(i_2-\floor{K/2})]\sigma_\nu^2$, representing the drift of $\nu_1$ and $\nu_2$ within the interval $[i_2-F\tc:i_2]$. Our observation of $\{\alpha_n\}$ is the sequence of $\bar \alpha_{i_2}$ in~\eqref{eq:2AP_phi_hat}, which follows approximately the model 
\begin{equation}
    \bar\alpha_{n} = \alpha_{n} + \xi_{n} + \mu_n \label{eq:kalman_observation}
\end{equation}
where $\xi_n \sim \normal(0,\sigma_\xi^2)$, with $\sigma_\xi^2 = 2(i_1-\floor{K/2})\sigma_\nu^2$, is the noise due to the drift of $\nu_1$ and $\nu_2$ within the interval $[i_2]_{\floor{K/2}}:i_1]$, and $\mu_n \sim \normal\big(0,\Pdl^{-1}\opnorm{\matG}^{-2}\big)$ is the total measurement error of $\alpha\of{2\to 1}_{i_1}$ and~$\alpha\of{1\to 2}_{i_2}$. 
Notice that $\xi_n$ is contained in~$\zeta_n$ because  $[i_2]_{\floor{K/2}}:i_1] \subset [i_2-F\tc:i_2]$. Thus, the observation noise and process noise in~\eqref{eq:kalman_process} and~\eqref{eq:kalman_observation} are correlated. 
Specifically,
    $\Exop[\zeta_n(\xi_n + \mu_n)] = \Exop[\xi_n^2] = \sigma_\xi^2$.
Therefore, we apply a generalized Kalman filter~\cite[Sec.~7.1]{Simon2006optimal} that captures this correlation. 
In particular, the Kalman gain is computed as
\begin{equation}
    \kappa_n = \frac{P_{n-1} + \sigma_\xi^2}{P_{n-1} + 3\sigma_\xi^2  + \Pdl^{-1}\opnorm{\matG}^{-2}}
\end{equation}
where $P_n$ is the current error variance. The phase estimate is then updated as 
\begin{equation}
    \hat\alpha_{n} =  \hat\alpha_{n-1} + \kappa_n {\rm wrap}(\bar\alpha_n - \hat\alpha_{n-1} )
\end{equation}
where the wrapping function ${\rm wrap}(\alpha) = [(\alpha + \pi) \mod 2\pi] - \pi$ is used to resolve the circular nonlinearity~\cite{Markovic16}. The error variance is updated as 
\begin{equation}
    P_{n}  = P_{n-1} - \kappa_n (P_{n-1} + \sigma_\xi^2) + \sigma_\zeta^2.
\end{equation}
The current filter output $\hat\alpha_{n}$ is used to estimate~$\phi\of{2}_i - \phi\of{1}_i$.

We can verify that $\kappa_n$ becomes closer to $1$ if $\Pdl\opnorm{\matG}^2$ increases. That is, the Kalman filter output becomes closer to the direct measurement if the inter-\gls{AP} receive \gls{SNR} is improved, in which case the quality of the phase measurements is high. We shall verify this effect in Section~\ref{sec:result}.

\section{Achievable Spectral Efficiency} \label{sec:rate}

We derive the achievable downlink \gls{SE} 
after phase synchronization. Let $\psi_{k,i}$ be the phase compensation term of \gls{UE}~$k$ at sample $i$. That is, $\psi_{k,i}$ is \gls{UE}~$k$'s current estimate of $\phi_i\of{1}$ (recall~\eqref{eq:manyAP_theta}). Following~\cite[Chap.~3]{marzetta2016fundamentals}, we write the received signals as a deterministic gain times the signal of interest plus uncorrelated noise that comprises beamforming gain uncertainty, multi-user interference, and \gls{AWGN}. We then use the ``use and forget'' trick~\cite{marzetta2016fundamentals} to lower-bound the ergodic capacity, inherently assuming that every codeword spans many realizations of all randomness, including the phase calibration errors. 
Specifically, the effective received signal is written  
as
\begin{align}
    e^{\jmath \revise{\psi_{k,i}}} y_{k,i} &= \DS_{k,i} s_{k,i} + \BU_{k,i} s_{k,i} + \UI_{k,i} + e^{\jmath \revise{\psi_{k,i}}} z_{k,i}. \label{eq:rx_conventionalTDD}
\end{align}
where 
\begin{align}
    \DS_{k,i} &= \sqrt{\Pdl}\Exop\left[\sum_{\ell = 1}^2 a_{\ell,i} \sqrt{\frac{\eta_{k,\ell}}{N \gamma_{k,\ell}}} \Delta_{k,\ell,i} \tp{\vecq}_{k,\ell,i} \hat\vecq_{k,\ell,i}^*\right], \label{eq:tmp410} \\ 
    \!\!\!\BU_{k,i} &= \!\sqrt{\Pdl\!} \sum_{\ell = 1}^2 \!a_{\ell,i} \sqrt{\!\frac{\eta_{k,\ell}}{N \gamma_{k,\ell}}} \Delta_{k,\ell,i} \tp{\vecq}_{k,\ell,i} \hat\vecq_{k,\ell,i}^* \!- \DS_{k,i}, \! \label{eq:conj_BU}\\ 
    \UI_{k,i} &= \!\!\!\sum_{k'=1, k'\ne k}^K   \underbrace{\!\!\Bigg(\!\!\sqrt{\Pdl}\sum_{\ell = 1}^2 a_{\ell,i}  \sqrt{\frac{\eta_{k',\ell}}{N \gamma_{k',\ell}}}  \Delta_{k,\ell,i} \tp{\vecq}_{k,\ell,i} \hat\vecq_{k',\ell,i}^* \!\Bigg)\!}_{= \UI_{k,k',i}} \notag \\
    &\quad \cdot s_{k',i}. 
\end{align}
Here, the term
\begin{equation}
    \Delta_{k,\ell,i} = \exp[\jmath (-\nu_{\ell,i} -\nu_{\ell,[i]_k} + \revise{\theta_{\ell,i}} + \revise{\psi_{k,i}})]. \label{eq:Delta}
\end{equation}
represents residual multiplicative phase noise after compensation.
The terms $\DS_{k,i}$, $\BU_{k,i}$, and $\UI_{k,i}$ represent the strength of the desired signal (DS), the beamforming gain uncertainty (BU), and the (residual) interference, respectively.

\subsection{Achievable Rate}
Let $\setJ = \{j_1, j_2, \dots\}$ be a sequence of samples across slots such that $\{\Delta_{k,\ell,i}\}_{i \in \setJ}$ and $\{a_{\ell,i}\}_{i \in \setJ}$ are ergodic processes. For the process $\{a_{\ell,i}\}_{i \in \setJ}$, it suffices to have that $a_{\ell,i}$ remains constant (either $0$ or $1$) for $i \in \setJ$.
We derive the achievable rate when channel coding is performed across $i \in \setJ$. 

For $i\in \setJ$, we can easily verify that the effective noise, containing the last three terms in~\eqref{eq:rx_conventionalTDD}, is uncorrelated with the desired signal for both beamforming schemes. Therefore, using the \revise{``use and forget'' trick~\cite{marzetta2016fundamentals}} 
and the fact that uncorrelated Gaussian noise represents the worst case, we obtain an achievable rate for coding over the sample sequence $\setJ$, represented by time index~$i$, as 
    \begin{align}
        R_{k,i} = \log_2\left(1+\frac{\abs{\DS_{k,i}}^2}{\Exop\big[\abs{\BU_{k,i}}^2\big] + 
        \Exop\big[\abs{\UI_{k,i}}^2\big] + 1} \right). \label{eq:rate_i}
    \end{align}
In the next theorem, we provide an expression for this achievable rate where the expectations are computed in closed form except for the randomness of $\Delta_{k,\ell,i}$. 

\begin{thm}
\label{th:rate_i}
    An achievable rate of the downlink transmission from the \glspl{AP} to user~$k$ when channel coding is applied across the sequence of samples $\setJ$, represented by time index~$i$, is given by~\eqref{eq:rate_i_conj}, shown at the top of the next page. 
    \begin{figure*}[t]
        \begin{align}
            R_{k,i} &= 
            \log_2\Bigg(1+\frac{N \Pdl\abs{ \sum_{\ell = 1}^2 a_{\ell,i} \sqrt{\eta_{k,\ell} \gamma_{k,\ell}} \Exop\left[\Delta_{k,\ell,i}\right]}^2}{N {\Pdl}\sum_{\ell = 1}^2 a_{\ell,i} \eta_{k,\ell} \gamma_{k,\ell} \big(1 - \left|\Exop\left[\Delta_{k,\ell,i}\right]\right|^2\big) +  \Pdl  \sum_{\ell = 1}^2 a_{\ell,i} \beta_{k,\ell} \sum_{k' = 1}^K \eta_{k',\ell}   +  1} \Bigg), \label{eq:rate_i_conj} 
        \end{align}
        \hrule
        \vspace{-.3cm}
    \end{figure*}
    
\end{thm}
\begin{proof}
    See Appendix~\ref{proof:rate_i}.
\end{proof}


    Comparing~\eqref{eq:rate_i_conj} with the no-phase-noise counterparts in~\cite[Eq.~(24)]{Ngo17_CF}. 
    we 
        see that phase noise leads to the down-scaling factor $\Exop[\Delta_{k,\ell,i}]$ in the DS strength (the numerator in~\eqref{eq:rate_i_conj}), and a new BU term  (the first term in the denominator in~\eqref{eq:rate_i_conj}). 
 %
        %
        Furthermore, phase noise does not affect the strength of inter-user interference (the second term in the denominators in~\eqref{eq:rate_i_conj}).
        %
        It is obvious that $|\Exop[\Delta_{\ell,i,k}]| \le 1$, with equality achieved when phase noise is not present or is perfectly compensated for. 
        Imperfect compensation of phase noise makes $|\Exop[\Delta_{\ell,i,k}]|$ strictly smaller than $1$, and thus decreases the rate.

\vspace{-.05cm}
\subsection{Achievable Spectral Efficiency}
We divide the time horizon into frames of $F$ slots in the same manner as in Section~\ref{sec:shifted_TDD}. Let $i_{f,1}, \dots, i_{f,F \tc}$ be the indices of the samples in frame~$f$. Each cross-frame sequence $\setI_n = \{i_{1,n}, i_{2,n}, \dots\}$, $n\in [F \tc]$, satisfy the assumptions of the sequence $\setJ$ described in the previous subsection. Specifically, for each $n\in[F\tc]$, the process $\{\Delta_{k,\ell,i}\}_{i \in \setI_n}$ contains i) the phase drift during the offset between time $i$ and the latest time when $\theta_{\ell,i}$ and $\psi_{k,i}$ are reset, and ii) the estimation error of the phase components used to compute $\theta_{\ell,i}$ and $\psi_{k,i}$. Both processes can be verified to be ergodic.
We apply multiple channel codes, one for each of the $F \tc$ sequences $\setI_n$, $n\in[F \tc]$, and leverage Theorem~\ref{th:rate_i} to obtain the following achievable \gls{SE}. 
\begin{cor}
    An achievable downlink \gls{SE} of \gls{UE}~$k$ is
    \begin{equation}
        \SEdl_k = \frac{1}{F \tc} \sum_{n=1}^{F \tc} R_{k,n} \quad \text{bit/s/Hz}
    \end{equation}
    where 
    $R_{k,n}$ is the achievable rate for the sequence $\setI_n$, computed as in~\eqref{eq:rate_i_conj}. 
\end{cor}

\vspace{-.25cm}
\section{Numerical Experiments}
\label{sec:result}

We numerically evaluate the achievable downlink \gls{SE} of the proposed broken \gls{TDD} flow. We consider a setting with $N = 64$ antennas per \gls{AP}, $K = 10$ \glspl{UE}, carrier frequency $f\sub{c} = 2\GHz$ and \revise{signal bandwidth $f\sub{s} = 20\MHz$.} Each coherence block has length $\tc = 100$ samples, 
including two periods of $\tg = 3$ guard samples, $\tpl = 10$ uplink pilot samples, and $\td = \frac12(\tc \!-\! \tpl \!-\! 2\tg) = 42$ downlink data samples. To focus on the impact of phase noise, we consider a static setting where $\beta_{k,\ell} = -20\dB$ for every pair of \gls{AP}~$\ell$ and \gls{UE}~$k$. Each \gls{AP} allocates power equally across the \glspl{UE}, i.e., $\eta_{k,\ell} = 1/K$, $\forall k,\ell$. We also let $\matG$ have \gls{iid} $\jpg(0,\beta_G)$ entries and define the \gls{SNR} between the \glspl{AP} as $\SNR\sub{AP} = \Pdl \beta_G$. 
The power constraints of the \glspl{UE} and \gls{AP} are given by $\Ppl = 20\dB$ and $\Pdl = 2\Ppl$, respectively.
A downlink demodulation pilot (to estimate $c$) is transmitted at the first sample of the downlink period in each slot.
\revise{We consider two types of oscillators with $c_{\nu} = 5\times 10^{-18}$ and $c_{\nu} = 1.58\times 10^{-17}$, corresponding to phase noise spectrum levels $-90\dBcHz$ and $-85\dBcHz$, respectively,\footnote{\revise{Typical values are between $-120\dBcHz$ and $-80\dBcHz$~\cite[Tab.~1]{Piemontese2024}.}} at $100\kHz$ offset~\cite{Piemontese2024}.} 
The expectation $\Exop[\Delta_{k,\ell,i}]$ is computed over $10^4$ realizations of consecutive $F$-slot frames. 

In Fig.~\ref{fig:LO1}, we show the average per-\gls{UE} downlink \gls{SE} as a function of $F$ for $c_{\nu} = 5\times 10^{-18}$ and $\SNR\sub{AP} \in \{-20\dB, \revise{-10\dB}\}$. 
Phase compensation is done with the direct estimation $\bar\alpha_{i_2}$ in~\eqref{eq:2AP_phi_hat} or the Kalman filter output. 
\revise{We see that the Kalman filter helps improve significantly the \gls{SE} with respect to direct phase estimation, especially if the inter-AP link is weak, i.e., $\SNR\sub{AP} = -20\dB$. 
Furthermore, it is optimal to re-estimate the phase after $2$ slots in this setting.  
We also show a baseline with AP~$2$ turned off, corresponding to $\ts = 0$ and $a_{2,i} = 0, \forall i$. In this case, AP~1 still transmits demodulation pilots but no inter-AP phase synchronization is needed. The \gls{SE} achieved with this baseline is largely below schemes with phase-calibrated coherent beamforming from both \glspl{AP}.}
\begin{figure}
    \centering
    \begin{tikzpicture}
        \begin{axis}[
            width=3.5in,
            height=2.3in,
            grid=both,
            xlabel={frame length, $F$ [slots]},
            ylabel={Average per-\gls{UE} \gls{SE} [bit/s/Hz]},
            legend style={at={(0.99,.99)}, anchor=north east,draw=none,text opacity=1,opacity = .8},
            legend cell align={left},
            line width=1pt,
            mark size=2pt,
            xmin=1, xmax=10,
            ymin=0.8, ymax=1.5,
            xtick={1,2,4,6,...,20},
            ytick={.8,.9,1,1.1,1.2,1.3,1.4,1.5},
            yticklabel style={/pgf/number format/fixed},
            ylabel style={yshift=-.2cm}
        ]
        
        \addplot[
            color=red,
        ]
        table[row sep=crcr]{
           1.0e+00 1.2354e+00 \\ 
2.0e+00 1.2517e+00 \\ 
3.0e+00 1.2466e+00 \\ 
4.0e+00 1.2351e+00 \\ 
5.0e+00 1.2236e+00 \\ 
6.0e+00 1.2148e+00 \\ 
7.0e+00 1.2096e+00 \\ 
8.0e+00 1.1967e+00 \\ 
9.0e+00 1.1920e+00 \\ 
1.0e+01 1.1764e+00 \\ 
        };
        \addlegendentry{\small with Kalman filter}
    
        \addplot[dashed,
            color=blue,
            mark=square*, mark options={solid, fill=white}
        ]
        table[row sep=crcr]{
            1.0e+00 1.1790e+00 \\ 
2.0e+00 1.2149e+00 \\ 
3.0e+00 1.2109e+00 \\ 
4.0e+00 1.2043e+00 \\ 
5.0e+00 1.1901e+00 \\ 
6.0e+00 1.1878e+00 \\ 
7.0e+00 1.1866e+00 \\ 
8.0e+00 1.1749e+00 \\ 
9.0e+00 1.1722e+00 \\ 
1.0e+01 1.1574e+00 \\ 
        };
        \addlegendentry{\small with direct phase estimation}

        \addplot[dashdotted,
            color=black
        ]
        table[row sep=crcr]{
            1.0e+00 8.4436e-01 \\ 
2.0e+00 8.4385e-01 \\ 
3.0e+00 8.4351e-01 \\ 
4.0e+00 8.4403e-01 \\ 
5.0e+00 8.4389e-01 \\ 
6.0e+00 8.4391e-01 \\ 
7.0e+00 8.4369e-01 \\ 
8.0e+00 8.4356e-01 \\ 
9.0e+00 8.4397e-01 \\ 
1.0e+01 8.4373e-01 \\ 
        };
        \addlegendentry{\small with only AP~$1$}
        
        \addplot[
            color=red,
        ]
        table[row sep=crcr]{
            1.0e+00 1.1817e+00 \\ 
2.0e+00 1.1869e+00 \\ 
3.0e+00 1.1712e+00 \\ 
4.0e+00 1.1544e+00 \\ 
5.0e+00 1.1405e+00 \\ 
6.0e+00 1.1145e+00 \\ 
7.0e+00 1.1007e+00 \\ 
8.0e+00 1.0799e+00 \\ 
9.0e+00 1.0710e+00 \\ 
1.0e+01 1.0593e+00 \\
        };
    
        \addplot[dashed,
            color=blue,
            mark=square*, mark options={solid, fill=white}
        ]
        table[row sep=crcr]{
            1.0e+00 9.6871e-01 \\ 
2.0e+00 1.0064e+00 \\ 
3.0e+00 1.0008e+00 \\ 
4.0e+00 9.9891e-01 \\ 
5.0e+00 9.9510e-01 \\ 
6.0e+00 9.8147e-01 \\ 
7.0e+00 9.7955e-01 \\ 
8.0e+00 9.7751e-01 \\ 
9.0e+00 9.7632e-01 \\ 
1.0e+01 9.6741e-01 \\ 
        };


        \draw[thin] (axis cs:3,1.235) ellipse [x radius=1.5mm, y radius=3mm];
        \node[align = center] at (axis cs:2.5,1.35) {$\SNR\sub{AP} =$ \\ $ -15\dB$};

        \draw[thin] (axis cs:8,1.025) ellipse [x radius=1.5mm, y radius=5mm];
        \node[align = center] at (axis cs:6,1.04) {$\SNR\sub{AP} = -20\dB$};
        

        
        \end{axis}
    \end{tikzpicture}
    \caption{Average per-\gls{UE} downlink \gls{SE} vs. the number of slots before the phase is re-estimated. We set $c_{\nu} = 5\times 10^{-18}$.}
    \label{fig:LO1}
    \vspace{-.3cm}
\end{figure}

\begin{figure}
    \centering
    \begin{tikzpicture}
        \begin{axis}[
            width=3.5in,
            height=2.3in,
            grid=both,
            xlabel={frame length, $F$ [slots]},
            ylabel={Average per-\gls{UE} \gls{SE} [bit/s/Hz]},
            legend style={at={(0.99,0.99)}, anchor=north east,draw=none},
            legend cell align={left},
            line width=1pt,
            mark size=2pt,
            xmin=1, xmax=10,
            ymin=0.8, ymax=1.5,
            xtick={1,2,4,6,...,20},
            ytick={.8,.9,1,1.1,1.2,1.3,1.4,1.5},
            yticklabel style={/pgf/number format/fixed},
            ylabel style={yshift=-.2cm}
        ]

        
        \addplot[
            color=red,
        ]
        table[row sep=crcr]{
            1.0e+00 1.1212e+00 \\ 
2.0e+00 1.0833e+00 \\ 
3.0e+00 1.0573e+00 \\ 
4.0e+00 1.0381e+00 \\ 
5.0e+00 1.0048e+00 \\ 
6.0e+00 9.7193e-01 \\ 
7.0e+00 9.4898e-01 \\ 
8.0e+00 9.2732e-01 \\ 
9.0e+00 9.1001e-01 \\ 
1.0e+01 9.0035e-01 \\ 
        };
        \addlegendentry{\small with Kalman filter}
    
        \addplot[dashed,
            color=blue,
            mark=square*, mark options={solid, fill=white}
        ]
        table[row sep=crcr]{
            1.0e+00 9.5578e-01 \\ 
2.0e+00 9.6941e-01 \\ 
3.0e+00 9.5730e-01 \\ 
4.0e+00 9.6179e-01 \\ 
5.0e+00 9.4348e-01 \\ 
6.0e+00 9.2373e-01 \\ 
7.0e+00 9.0792e-01 \\ 
8.0e+00 8.9245e-01 \\ 
9.0e+00 8.7860e-01 \\ 
1.0e+01 8.6789e-01 \\ 
        };
        \addlegendentry{\small with direct phase estimation}

         \addplot[dashdotted,
            color=black
        ]
        table[row sep=crcr]{
            1.0e+00 8.4021e-01 \\ 
2.0e+00 8.4066e-01 \\ 
3.0e+00 8.4064e-01 \\ 
4.0e+00 8.4052e-01 \\ 
5.0e+00 8.4036e-01 \\ 
6.0e+00 8.4065e-01 \\ 
7.0e+00 8.4035e-01 \\ 
8.0e+00 8.4058e-01 \\ 
9.0e+00 8.4059e-01 \\ 
1.0e+01 8.4030e-01 \\
        };
        \addlegendentry{\small with only AP~$1$}
        
        \addplot[
            color=red,
        ]
        table[row sep=crcr]{
            1.0e+00 1.1931e+00 \\ 
2.0e+00 1.1922e+00 \\ 
3.0e+00 1.1665e+00 \\ 
4.0e+00 1.1474e+00 \\ 
5.0e+00 1.1252e+00 \\ 
6.0e+00 1.1036e+00 \\ 
7.0e+00 1.0854e+00 \\ 
8.0e+00 1.0555e+00 \\ 
9.0e+00 1.0422e+00 \\ 
1.0e+01 1.0208e+00 \\ 
        };
    
        \addplot[dashed,
            color=blue,
            mark=square*, mark options={solid, fill=white}
        ]
        table[row sep=crcr]{
        1.0e+00 1.1616e+00 \\ 
2.0e+00 1.1729e+00 \\ 
3.0e+00 1.1502e+00 \\ 
4.0e+00 1.1320e+00 \\ 
5.0e+00 1.1167e+00 \\ 
6.0e+00 1.0942e+00 \\ 
7.0e+00 1.0806e+00 \\ 
8.0e+00 1.0499e+00 \\ 
9.0e+00 1.0348e+00 \\ 
1.0e+01 1.0166e+00 \\ 
        };

        \draw[thin] (axis cs:5,1.12) ellipse [x radius=1.5mm, y radius=3mm];
        \node[align = center] at (axis cs:6.9,1.16) {$\SNR\sub{AP} = -15\dB$};

        \draw[thin] (axis cs:5,.97) ellipse [x radius=1.5mm, y radius=4mm];
        \node[align = center] at (axis cs:3.05,.91) {$\SNR\sub{AP} = -20\dB$};
        

        
        \end{axis}
    \end{tikzpicture}
    \caption{Similar to Fig.~\ref{fig:LO2} but with lower-quality \glspl{LO}, i.e., $c_{\nu} = 1.58\times 10^{-17}$.}
    \label{fig:LO2}
    \vspace{-.3cm}
\end{figure}

\revise{In Fig.~\ref{fig:LO2}, 
we consider lower-quality \glspl{LO} with $c_{\nu} = 1.58\times 10^{-17}$. 
Similar observations as in Fig.~\ref{fig:LO2} hold, except that the achievable \gls{SE} decreases more quickly as $F$ grows. With Kalman filter, it is optimal to re-estimate the phase shift in every slot. 
Furthermore, we see that the gain from using the Kalman filter is less significant in this setting than in Fig.~\ref{fig:LO1}.}



\section{Conclusions} \label{sec:conclusion}
We proposed a mechanism to break the \gls{TDD} flow in distributed antenna systems so that phase synchronization can be performed using bi-directional phase measurements between the \glspl{AP}. We also designed a Kalman filter for phase tracking based on these measurements. Our results highlighted i) the need for frequent re-estimation of the inter-\gls{AP} phase disparity for coherent beamforming when the \glspl{LO} drift rapidly, and ii) the advantage of using the Kalman filter when the inter-\gls{AP} link is weak \revise{or the \glspl{LO} drift slowly}. 
In future work, we will consider a setting with many \glspl{AP} and address \gls{AP} scheduling to determine which \glspl{AP} should transmit synchronization signals and which should receive in a given slot. 

\begin{appendices}

\section{Proof of Theorem~\ref{th:rate_i}} \label{proof:rate_i}

In the following, we omit the subscript $i$ in $\vecq_{k,\ell,i}$ and $\hat \vecq_{k,\ell,i}$ for notational simplicity.
We will use the following lemma. 

\begin{lem} \label{lem:h_estimate_property}
    It holds that 
    \begin{enumerate}
        \item $\hat \vecq_{k,\ell} \sim \jpg\left(\veczero, \gamma_{k,\ell}\matidentity_N\right)$  with $\gamma_{k,\ell} = \sqrt{\Ppl K} \beta_{k,\ell} c_{k,\ell}$;
        
        \item $\Exop[\tp{\tilde \vecq}_{k,\ell} \hat \vecq_{k,\ell}^* \given \nu_{\ell,[i]_k}] = 0$ where $\tilde \vecq_{k,\ell} =  \vecq_{k,\ell} - \hat \vecq_{k,\ell} \sim  \jpg\left(\veczero, \beta_{k,\ell} - \gamma_{k,\ell}\matidentity_N\right)$ is the channel estimation error;

        
        \item $\Exop[|\tp{\tilde \vecq}_{k,\ell} \hat\vecq_{k,\ell}^*|^2] = N \gamma_{k,\ell}(\beta_{k,\ell} - \gamma_{k,\ell})$;
        
        \item $\Exop[\|\hat\vecq_{k,\ell}\|^4] = N(N + 1)\gamma_{k,\ell}^2$.
    \end{enumerate}
    
\end{lem}
\begin{proof}
    Conditioned on $\nu_{\ell,[i]_k}$, $\hat{\vecq}_{k,\ell}$ is the \gls{MMSE} estimate of $\vecq_{k,\ell}$. The lemma follows directly from properties of the \gls{MMSE} estimate and some simple manipulations.
\end{proof}


The derivation of the achievable rate follows similar steps as in~\cite[Chap.~3]{marzetta2016fundamentals} and~\cite{Ngo17_CF}, but with the extra factor $\Delta_{k,\ell,i}$.

\subsection{Conjugate Beamforming}

\subsubsection{Compute $\DS_{k,i}$} We proceed as
\begin{align}
    &\DS_{k,i} = \sqrt{\Pdl} \sum_{\ell = 1}^2 a_{\ell,i} \sqrt{\frac{\eta_{k,\ell}}{N \gamma_{k,\ell}}} \Exop\left[\Delta_{k,\ell,i} \tp{\vecq_{k,\ell}} \hat\vecq_{k,\ell}^* \right] \\
    &= \sqrt{\Pdl} \sum_{\ell = 1}^2  \!a_{\ell,i} \sqrt{\frac{\eta_{k,\ell}}{N \gamma_{k,\ell}}} \Exop\!\Big[\Delta_{k,\ell,i} \Exop\!\big[\tp{\vecq_{k,\ell}} \hat\vecq_{k,\ell}^* \given \nu_{\ell,[i]_k}\big] \Big]  \label{eq:tmp258}\\
    &= \sqrt{\Pdl} \sum_{\ell = 1}^2 a_{\ell,i} \sqrt{\eta_{k,\ell} N \gamma_{k,\ell}} \Exop\left[\Delta_{k,\ell,i}\right], \label{eq:DS}
\end{align} 
where \eqref{eq:tmp258} holds because given $\nu_{\ell,[i]_k}$, $\tp{\vecq_{k,\ell}} \hat\vecq_{k,\ell}^*$ is independent of $\Delta_{k,\ell,i}$, and \eqref{eq:DS} follows from $\Exop\left[\tp{\vecq_{k,\ell}} \hat\vecq_{k,\ell}^* \given \nu_{\ell,[i]_k} \right] = \Exop\left[\|\hat\vecq_{k,\ell}\|^2 \given \nu_{\ell,[i]_k} \right] = N \gamma_{k,\ell}$.

\subsubsection{Compute $\Exop\left[\abs{\BU_{k,i}}^2\right]$} 
Notice that the variables $\Delta_{k,\ell,i} \tp{\vecq}_{k,\ell} \hat\vecq_{k,\ell}^*$ are uncorrelated across $\ell$.
Using the fact that the variance of the sum of uncorrelated random variables is equal to the sum of the variances, we have that 
\begin{align}
    &\Exop\left[\abs{\BU_{k,i}}^2\right] \notag \\
    &= {\Pdl}\!\sum_{\ell = 1}^2 a_{\ell,i} \frac{\eta_{k,\ell}}{N \gamma_{k,\ell}\!} \Exop\!\left[\left|\Delta_{k,\ell,i} \tp{\vecq}_{k,\ell} \hat\vecq_{k,\ell}^* - \Exop\!\left[\Delta_{k,\ell,i} \tp{\vecq}_{k,\ell} \hat\vecq_{k,\ell}^*\right] \right|^2\right] \\ 
    &= {\Pdl}\sum_{\ell = 1}^2 a_{\ell,i} \frac{\eta_{k,\ell}}{N \gamma_{k,\ell}} \Big(\underbrace{\Exop\left[\left|\Delta_{k,\ell,i} \tp{\vecq}_{k,\ell} \hat\vecq_{k,\ell}^*\right|^2\right]}_{A} \notag \\
    &\qquad - \underbrace{\left|\Exop\left[\Delta_{k,\ell,i} \tp{\vecq}_{k,\ell} \hat\vecq_{k,\ell}^*\right] \right|^2}_{B} \Big). \label{eq:tmpBU}
\end{align}
The same computation leading to~\eqref{eq:DS} shows that $B =~\left|\Exop\left[\Delta_{k,\ell,i}\right]\right|^2 N^2 \gamma_{k,\ell}^2$. We compute the term $A$ as follows
\begin{align}
    A &= \Exop_{\nu_{\ell,[i]_k}}\left[\Exop\left[\left|\tp{\tilde \vecq}_{k,\ell} \hat\vecq_{k,\ell}^* +  \|\hat\vecq_{k,\ell}\|^2\right|^2 \given \nu_{\ell,[i]_k}\right]\right] \\ 
    &= \Exop\left[\left|\tp{\tilde \vecq}_{k,\ell} \hat\vecq_{k,\ell}^*\right|^2\right] +  \Exop\left[\|\hat\vecq_{k,\ell}\|^4\right] \label{eq:tmp280} \\ 
    &= N \gamma_{k,\ell} (\beta_{k,\ell} - \gamma_{k,\ell}) + N(N + 1) \gamma_{k,\ell}^2 \label{eq:tmp293} \\ 
    &= N \gamma_{k,\ell} \beta_{k,\ell} + N^2 \gamma_{k,\ell}^2,
\end{align}
where~\eqref{eq:tmp280} follows since given $\nu_{\ell,[i]_k}$, $\tilde \vecq_{k,\ell}$ has mean $\veczero$ and is uncorelated with $\hat\vecq_{k,\ell}$, and~\eqref{eq:tmp293} follows from Lemma~\ref{lem:h_estimate_property}.
Using the computation of $A$ and $B$ in~\eqref{eq:tmpBU}, we obtain
\begin{multline}
    \Exop\left[\abs{\BU_{k,i}}^2\right] = \\ {\Pdl}\sum_{\ell = 1}^2 a_{\ell,i} \eta_{k,\ell} \big( \beta_{k,\ell} 
    + N \gamma_{k,\ell} \big(1 - \left|\Exop\left[\Delta_{k,\ell,i}\right]\right|^2\big) \big). \label{eq:BU}
\end{multline} 

\subsubsection{Compute $\Exop\left[\abs{\UI_{k,i}}^2\right]$} As the variables $\UI_{k,k',i}$ are uncoorelated across $k'$, we have that $ \Exop\left[\abs{\UI_{k,i}}^2\right] = \sum_{k' = 1, k'\ne k}^K \Exop\left[\abs{\UI_{k,k',i}}^2\right]$. Next, expanding $\hat\vecq_{k',\ell}^*$, we obtain
\begin{align}
    &\Exop\left[\abs{\UI_{k,k',i}}^2\right]
    \notag \\
    &= \Exop\Bigg[\bigg|\sqrt{\Pdl} \sum_{\ell = 1}^2 a_{\ell,i}  \sqrt{\frac{\eta_{k',\ell}}{N \gamma_{k',\ell}}} \Delta_{k,\ell,i} \notag \\
    & \qquad \cdot \tp{\vecq}_{k,\ell} c_{k',\ell}\Big(\sqrt{\Ppl K} \vecq_{k',\ell} + \vecz\supp{pilot}_{k',\ell}\Big)\bigg|^2\Bigg] \\ 
    &= \Pdl \Exop\!\Bigg[\bigg|\sum_{\ell = 1}^2 a_{\ell,i} \sqrt{\frac{\eta_{k',\ell}}{N \gamma_{k',\ell}\!}} \Delta_{k,\ell,i} c_{k',\ell}\sqrt{\Ppl K}  
    \tp{\vecq}_{k,\ell}  \vecq_{k',\ell}\bigg|^2\Bigg] \notag \\
    &\quad + \Pdl \Exop\!\Bigg[\bigg|\sum_{\ell = 1}^2 a_{\ell,i}  \sqrt{\frac{\eta_{k',\ell}}{N \gamma_{k',\ell}}} \Delta_{k,\ell,i} c_{k',\ell} 
    \tp{\vecq}_{k,\ell} \vecz\supp{pilot}_{k',\ell}\bigg|^2\Bigg] \label{eq:tmp309} \\ 
    &= \Pdl \sum_{\ell = 1}^2 a_{\ell,i}  \frac{\eta_{k',\ell}}{N \gamma_{k',\ell}} c^2_{k',\ell} \Ppl K \Exop\left[\abs{\Delta_{k,\ell,i} \tp{\vecq}_{k,\ell}  \vecq_{k',\ell}}^2\right] \notag \\
    &\quad + \Pdl \sum_{\ell = 1}^2 a_{\ell,i}  \frac{\eta_{k',\ell}}{N \gamma_{k',\ell}} c^2_{k',\ell}  \Exop\left[\abs{\Delta_{k,\ell,i} \tp{\vecq}_{k,\ell} \vecz\supp{pilot}_{k',\ell}}^2\right] \label{eq:tmp310} \\ 
    &= \Pdl \sum_{\ell = 1}^2 a_{\ell,i}   \frac{\eta_{k',\ell}}{N \gamma_{k',\ell}} c^2_{k',\ell}  \Big(\Ppl K\Exop\left[\abs{\tp{\vech}_{k,\ell}  \vech_{k',\ell}}^2\right] \notag \\
    &\qquad + \Exop\left[\big|\tp{\vech}_{k,\ell} \vecz\supp{pilot}_{k',\ell}\big|^2\right] \Big)\label{eq:tmp337} \\
    &= \Pdl \sum_{\ell = 1}^2 a_{\ell,i} \frac{\eta_{k',\ell}}{N \gamma_{k',\ell}} c^2_{k',\ell} N \beta_{k,\ell} \Big(\Ppl K  \beta_{k',\ell} + 1 \Big) \\
    &= \Pdl \sum_{\ell = 1}^2 a_{\ell,i} \eta_{k',\ell} \beta_{k,\ell},
    \label{eq:UI}
\end{align}
where~\eqref{eq:tmp309} holds because $\vecz\supp{pilot}_{k',\ell}$ is independent of $\vecq_{k',\ell}$ and has zero mean, ~\eqref{eq:tmp310} holds because both $\Delta_{k,\ell,i} \tp{\vecq}_{k,\ell}  \vecq_{k',\ell}$ and $\Delta_{k,\ell,i} \tp{\vecq}_{k,\ell} \vecz_{k',\ell}$ are uncorrelated across $\ell$ and have zero mean. 

Finally, substituting~\eqref{eq:DS}, \eqref{eq:BU}, and~\eqref{eq:UI} into~\eqref{eq:rate_i}, we obtain the expression~\eqref{eq:rate_i_conj} of the achievable rate $R_{k,i}$.

\end{appendices}

\bibliographystyle{IEEEtran}
\bibliography{IEEEabrv,./ref}

\end{document}